# Digital Deception: Generative Artificial Intelligence in Social Engineering and Phishing

MARC SCHMITT

Department of Computer Science, University of Oxford, UK

IVAN FLECHAIS

Department of Computer Science, University of Oxford, UK

**Abstract**

The advancement of Artificial Intelligence (AI) and Machine Learning (ML) has profound implications for both the utility and security of our digital interactions. This paper investigates the transformative role of Generative AI in Social Engineering (SE) attacks. We conduct a systematic review of social engineering and AI capabilities and use a theory of social engineering to identify three pillars where Generative AI amplifies the impact of SE attacks: Realistic Content Creation, Advanced Targeting and Personalization, and Automated Attack Infrastructure. We integrate these elements into a conceptual model designed to investigate the complex nature of AI-driven SE attacks—the Generative AI Social Engineering Framework. We further explore human implications and potential countermeasures to mitigate these risks. Our study aims to foster a deeper understanding of the risks, human implications, and countermeasures associated with this emerging paradigm, thereby contributing to a more secure and trustworthy human-computer interaction.

## 1 INTRODUCTION

Digital deception, fueled by the advancements in generative AI, poses a serious threat to our interconnected society. As AI systems become increasingly adept at mimicking human communication and trust signals, they present a new frontier for social engineering and phishing attacks. Given that social engineering attacks can arise in all types of interaction—computer-to-computer, human-to-computer, or human-to-human—urgent research is needed: both to understand the challenges, and to identify ways of protecting safe and meaningful digital interactions when such technologies are used for deceit.

The rapid advancements in Artificial Intelligence (AI) and Machine Learning (ML) are poised to revolutionize many aspects of our lives [1,2], with significant implications for the security and privacy of our digital interactions [3]. The growing dependence on technology opens a wide playground for cybercriminals, posing threats to all stakeholders in the global economy. Cybercrime is quickly increasing in scale and impact, causes serious financial harm, breaches individual privacy, and has the potential to break down critical infrastructure. The World Economic Forum [4] consistently ranks cyber threats as one of the most significant global risks. AI/ML technologies have demonstrated their potential for helping to detect social engineering and phishing attacks [5], however these technologies can also be applied in a malicious manner to amplify the capabilities and effectiveness of such attacks. Understanding the nature of AI-driven social engineering attacks and their techniques is crucial for individuals and organizations to implement proactive strategies, and defenses against this "relevant" and "permanent" cybersecurity threat [3,5].

The increasing sophistication of AI algorithms and the availability of vast amounts of data can empower malicious actors to develop more sophisticated and convincing phishing techniques. AI/ML can be misused to automate and personalize phishing attacks, generate persuasive content that mimics legitimate communication, and evade detection by traditional security measures [3,6,7]. As security threats evolve, it becomes crucial to review and explore the potential risks and vulnerabilities associated with AI/ML to better understand this new threat landscape, and to develop effective countermeasures for individuals, organizations, and broader digital ecosystems.

The recent rise of generative AI – in particular, large language models [8,9] – is a new emergent capability with the potential to carry out highly targeted social engineering campaigns at an industrial scale [10]. With the help of AI-enabled autonomous agents [11], it is possible to automate parts – potentially even the entire lifecycle – of a social engineering attack. In addition, fully trained SE models can theoretically learn from each attack and gradually improve their success rate. The emergence of advanced open-source generative AI tools represents significant progress in machine learning and AI, particularly in their capacity to produce sophisticated text, voice, graphics, and video content. This is particularly evident by recent developments in large language models (LLMs) such as GPT-4 [8] and PaLM 2 [9], which power the commercial chatbots ChatGPT and Bard, respectively. Those tools have raised substantial concerns about their potential misuse in social engineering attacks [10]. In conjunction with techniques like jailbreaking, prompt injection, and reverse psychology, these models pose a notable threat to the cybersecurity landscape [12], however much is still unknown about this threat and its characteristics. To address this gap, our research aims to investigate the following question: How can Generative AI amplify the effectiveness – and hence the global problem – of social engineering and phishing attacks?

This research aims to examine and analyze the various attack options available in social engineering and phishing, and identify those areas which are facilitated using AI/ML. It seeks to understand how threat actors can harness AI/ML to carry out sophisticated social engineering and phishing campaigns and to provide insights for threat intelligence to develop future mitigation strategies. The findings of this study will be valuable for researchers, practitioners, and policymakers, guiding them in understanding the evolving landscape of AI-enabled social engineering and phishing attacks.

This study addresses three research questions:

- RQ1: How can generative AI likely be used – by cybercriminals – to enhance the effectiveness of social engineering and phishing attacks?
- RQ2: How does the integration of AI in social engineering and phishing attacks – especially generative models – pose new challenges and demand the development of advanced defense mechanisms?
- RQ3: How can we effectively detect and mitigate these AI-driven attacks?

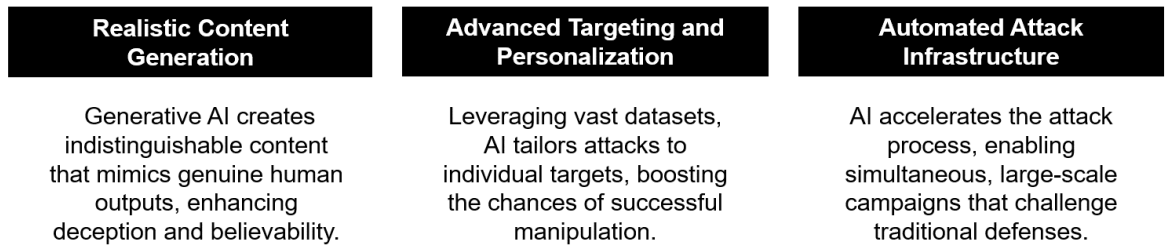

Figure 1: Three Pillar Framework of Generative AI-enabled Social Engineering Attacks

This paper is structured as follows: The methodology section outlines our research methods and shows the concrete steps taken to develop the Generative AI Social Engineering Framework. This is followed by a theoretical foundation section that discusses social engineering, phishing, and artificial intelligence capabilities. The section on generative AI-enabled social engineering attacks is divided into the three subsections: (i) realistic content generation, (ii) advanced targeting and personalization, and (iii) automated attack infrastructure, which are the three pillars of our framework (see Figure 1). The paper then discusses the impact of increasing capabilities and cost-effectiveness of AI-enabled phishing attacks, explores human implications and potential countermeasures, and concludes with details of future research in this area.

## 2 METHODOLOGY

To address the complexities and challenges posed by the integration of Generative AI in Social Engineering (SE), we employ a mixed-methods research approach that combines qualitative and framework-based analyses. Our methodology comprises the following steps:

- Systematic Review: We start by conducting a systematic review of existing literature to understand the state of the art in both Social Engineering and Generative AI. This review provides the foundation upon which we built our subsequent analyses.
- AI Capabilities Analysis: Leveraging the Social Engineering Framework developed by Mouton et al., [13], we analyze the capabilities of AI technologies, placing a specific emphasis on Generative AI. This allows us to understand how these technologies fit into existing SE tactics and strategies.
- Identification of Pillars: Based on our analysis, we identify three primary pillars where Generative AI exacerbates the impact of SE attacks. These pillars are: (1) Realistic Content Creation, (2) Advanced Targeting and Personalization, and (3) Automated Attack Infrastructure.
- Conceptual Framework Development: Informed by our findings, we develop a conceptual framework that we refer to as the "Generative AI Social Engineering Framework." The model offers a blueprint for deeper and broader investigations, and the derivation of potential implications and solutions.
- Applications: We then validate and apply the developed framework to (1) investigate the impact of generative AI on different types of phishing attacks in terms of threat amplification and cost-effectiveness, and (2) to identify countermeasures that can mitigate the risks associated with each of the three pillars of the framework.

By adopting this methodology, we aim to offer a robust and multidimensional analysis of the challenges and solutions associated with the role of Generative AI in Social Engineering and lay the groundwork for further research.

## 3 LITERATURE BACKGROUND

### 3.1 Social Engineering and Phishing

Deception plays a central role in both social engineering and phishing tactics. In both cases, the attacker aims to manipulate or trick their target into taking certain actions or revealing confidential information by posing as a trustworthy entity. Social engineering refers to the activity of manipulating people into performing actions or divulging confidential information, usually through deceptive means. It exploits human psychology rather than relying on technological vulnerabilities. See Table 1 for an overview of different social engineering types.

Table 1: Different Social Engineering Attack Types

| Type | Explanation |
| --- | --- |
| Pretexting | Attacker creates a fabricated scenario to obtain information or access from a target. |
| Baiting | Attacker lures target with appealing offer to compromise their security (e.g., free software). |
| Tailgating | Attacker deceives security personnel or employees to gain physical access to a restricted area. |
| Quizzing | Attacker tricks the target into disclosing information under the guise of a survey or quiz. |
| Scareware | Attacker tricks target into buying fake antivirus, claiming their computer is infected. |
| Phishing | Deceptive emails. The most well-known and widespread form of social engineering. |

Phishing attacks are used by adversaries to impersonate legitimate entities and send deceptive emails or texts, often containing malware attachments or links to fraudulent websites [14,15]. Phishing attacks exploit human psychology, relying on social engineering techniques to trick victims into providing their confidential information or performing actions that compromise their security [16].

The primary objectives of these scams are to infect devices with malware, steal sensitive data like usernames and credit card details, seize control of digital accounts such as email and social media, or directly persuade recipients to perform financial transactions. Phishing as a strategy exploits user interface flaws and takes advantage of the fact that humans have difficulties verifying URLs and dynamic content of rendered HTML documents [14]. It is often the first point of access to a victim's device or account and can lead to further attacks, hence exploiting the trust developed between the compromised account and its contacts. Phishing attacks have been a pervasive and persistent threat in the digital landscape, impacting individuals, businesses, and even government entities. Those attacks have evolved, becoming increasingly sophisticated and difficult to detect. Attackers constantly adapt their techniques, leveraging advancements in technology to create more convincing and personalized phishing campaigns [10,15]. The consequences of phishing attacks can be severe, ranging from financial losses and identity theft to unauthorized access to sensitive data or compromise of entire systems. See Table 2 for an overview of different phishing types.

Table 2: Different Phishing Attack Types

| Phishing Types | Explanation |
| --- | --- |
| Email Phishing | The attacker sends fraudulent emails that appear to come from legitimate sources (e.g., banks, government agencies) to trick recipients into revealing sensitive information. |
| Spear Phishing | Like email phishing, but the attacker customizes the deceptive email for a specific individual or organization. |
| Smishing | The attacker sends deceptive text messages (SMS) to trick recipients into revealing sensitive information or clicking on malicious links. |

| Phishing Types | Explanation |
| --- | --- |
| Whaling | A specialized form of spear phishing where high-profile targets, like CEOs or CFOs, are the primary focus. The intent is typically to manipulate these individuals into authorizing large money transfers or revealing sensitive corporate data. |
| Pharming | The attacker redirects the target's web traffic to a fake website that looks identical to a legitimate one, tricking them into entering their credentials or other sensitive information. |
| Vishing | The attacker uses deceptive phone calls to trick the target into divulging confidential information. |
| Social-Media Phishing | The attacker uses social media platforms (e.g. Facebook or Instagram) to deceive their targets [17]. |

There are also other concepts, such as "catfishing", where someone creates a fake identity on a social networking site, often to deceive someone into a fraudulent romantic relationship. While this is not "phishing" and the end goal might not always be financial gain or direct access to sensitive information, it is another example of deception through social platforms. Regardless of the method or platform, the core principle remains: attackers exploit human psychology and trust to deceive individuals. When we refer to phishing, we mean all types of phishing as outlined in Table 2, and when we refer to social engineering, we mean all types of social engineering (including phishing) as shown in Table 1.

Based on an analysis of different social engineering attacks, Mouton et. al. [13] propose a model of social engineering attacks outlining several phases and characteristic actions. In Figure 2, we reproduce the high level model of the framework, which shows how a social engineering attack progresses through different phases (attack formulation, information gathering, preparation, developing a relationship, exploiting a relationship, and debrief). We use this framework in Section 4 to help structure our analysis of Gen-AI capabilities on social engineering attacks.

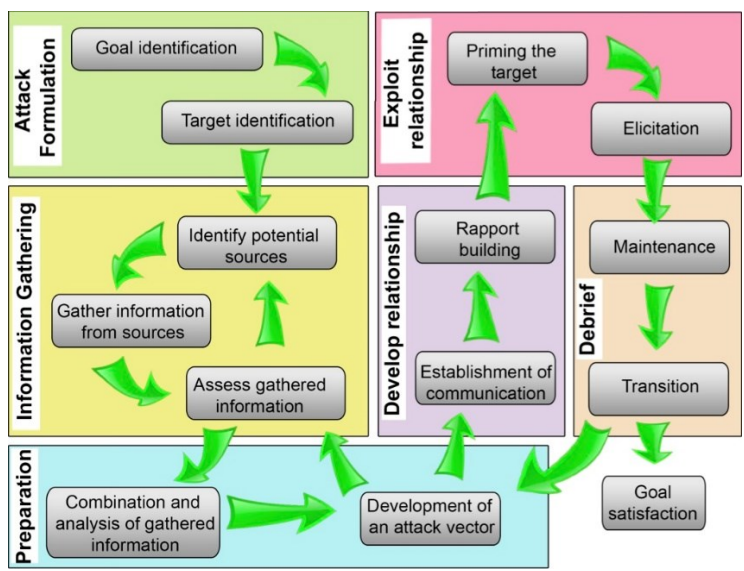

Figure 2: Social Engineering Attack Framework, reproduced from Mouton et. al. [13]

### 3.2 Artificial Intelligence Capabilities

Artificial Intelligence (AI) has emerged as a transformative technology, primarily driven by key factors such as the abundance of data, improved algorithms, and advances in hardware infrastructure. Machine Learning (ML), a subfield of AI, focuses on developing algorithms and models that can learn patterns and make predictions or decisions based on data. Machine learning can be divided into supervised learning, unsupervised learning, semi-supervised learning, and reinforcement learning [18,19]. Breakthroughs in deep learning, a subset of ML, have revolutionized the field by enabling the training of neural networks with multiple layers, leading to remarkable progress in areas like image recognition, natural language processing, and autonomous systems [20]. The impact of AI/ML is increasingly felt in various domains, including healthcare, finance, and cybersecurity [3,21,22]. AI-powered systems can automate tasks, enhance decision-making processes, and uncover valuable insights from vast amounts of data [18,23]. Generative AI has opened a novel frontier in the realm of AI. Generative Adversarial Networks (GANs), a pioneering approach within this domain, have become particularly noteworthy for their ability to produce incredibly realistic digital assets ranging from images to videos [24]. Recently, Large Language Models (LLMs) show the culmination of deep learning techniques applied to natural language processing. LLMs, such as GPT-4 and Palm 2 – which are based on the transformer architecture [25] – have shown unprecedented capabilities in generating realistic text content [8]. However, this power also raises ethical and security concerns, particularly as the line between human-generated and AI-generated content becomes increasingly blurred.

Table 3: Artificial Intelligence in a Nutshell

| Machine Learning | Description |
|---|---|
| Supervised Learning | A machine learning algorithm learns from labeled training data, where each example is paired with its corresponding target output. The algorithm generalizes from this labeled data to make predictions or classify new, unseen instances based on the learned patterns. |
| Unsupervised Learning | Involves learning patterns and structures from unlabeled data without explicit target outputs. The algorithm identifies inherent relationships, clusters, or patterns within the data to gain insights and make sense of the underlying information. |
| Semi-Supervised Learning | The algorithm learns from a combination of labeled and unlabeled data. By leveraging the unlabeled data along with the labeled data, the algorithm aims to improve its predictive performance and make use of the additional information available. |
| Reinforcement Learning | A learning paradigm where an agent learns to interact with an environment by taking actions and receiving feedback in the form of rewards or penalties. The agent aims to maximize its cumulative reward through an iterative process of exploration and exploitation. |
| Deep Learning | A human brain inspired subset of machine learning, uses neural networks with multiple layers to process and analyze large data sets. The ability to detect complex patterns enabled advances in image and voice recognition, natural language processing, and beyond. |
| Generative AI | AI models – based on deep learning – designed to generate new content that resembles the input data they have been trained on. Examples include Generative Adversarial Networks (GANs) that produce synthetic images and videos, and Large Language Models (LLMs) like GPT-4 and Palm 2, that generate realistic textual content. |

Machine learning has gradually introduced a huge basket of "AI capabilities" that can be harnessed for social engineering and phishing attacks. See Table 4 for an overview of existing AI capabilities and how they could potentially be applied in this context.

Table 4: Overview of Potential AI Capabilities in the Context of Social Engineering

| AI Capabilities | Explanation | ML Techniques |
|---|---|---|
| Generative AI | Generative AI involves algorithms that can generate content, such as text, images, or videos, based on patterns learned from existing data. In social engineering attacks, generative AI can create realistic and | Generative Adversarial Networks (GANs), Transformer models |

| AI Capabilities | Explanation | ML Techniques |
|---|---|---|
| | convincing attack vectors, such as phishing emails, by imitating human communication styles and context. | |
| AI Analysis | AI analysis refers to the application of machine learning and data analysis techniques to process and interpret data. In social engineering attacks, AI analysis can identify potential targets, assess their vulnerabilities, and predict their behavior based on patterns in gathered information. | Machine Learning, Classification and Regression, Natural Language Processing (NLP) |
| AI Scraping | AI scraping entails the use of automated tools, often driven by machine learning, to collect information from various online sources. In social engineering, AI scraping can swiftly gather data from social media profiles, public databases, and other sources to create detailed profiles of targets. | Web Scraping Libraries, Data Mining, Clustering Techniques |
| AI Automation | AI automation refers to the use of AI-driven systems to automate various tasks and processes. In social engineering, AI automation can initiate and maintain communication with targets, ensuring consistent interaction and reducing the risk of detection. | Rule-based Systems, Process Automation, Workflow Management |
| AI Chatbots | AI chatbots are computer programs that can simulate human conversation. In social engineering attacks, AI chatbots can engage targets in conversations to build trust, gather information, and manipulate emotions, all while emulating human-like interaction. | Large Language Models (LLMs), Contextual Chatbot Frameworks, Sequence-to-Sequence Models |
| AI Coordination | AI coordination involves the orchestration of tasks and interactions among different AI agents or components. In social engineering, AI coordination could ensure smooth transitions between different phases of the attack and maintains continuity, even if attackers change. | Multi-agent Systems, Coordination Algorithms, Task Allocation Methods |
| AI Assessment | AI assessment entails the use of algorithms to track, analyze, and evaluate the success of an attack. In social engineering, AI assessment can monitor the outcomes, such as compromised accounts or data leaks, to determine the effectiveness of the attack and refine future strategies. | Performance Metrics, Anomaly Detection, A/B Testing |

## 4 CAPABILITY ANALYSIS OF GENERATIVE AI IN SOCIAL ENGINEERING

Generative AI can be employed in social engineering and phishing scenarios in various ways, enhancing the effectiveness of these attacks by creating more convincing and targeted deceptive content. In this section, we will investigate the integration of AI capabilities into the various stages of the Social Engineering (SE) attack lifecycle introduced by Mouton et al. [13]. While Table 5 shows the entire SE attack framework and most machine learning capabilities, we will focus on the stages where the potential for Gen-AI is the largest.

Table 5: AI utilization for each stage of the social engineering attack lifecycle

| AI Utilization | Stage of SE Attack Lifecycle | Application | AI Capabilities |
|---|---|---|---|
| Attack Formulation | Goal Identification | Generating potential attack goals based on desired outcomes and vulnerabilities. | Generative AI |
| | Target Identification | Analyzing public data to identify potential targets based on roles and online presence. | AI Analysis |
| Information Gathering | Identify Potential Sources | Scraping potential sources of information from public sources and social media. | AI Scraping |

| AI Utilization | Stage of SE Attack Lifecycle | Application | AI Capabilities |
|---|---|---|---|
|  | Gather Information from Sources | Gathering relevant information from various data sources. | AI Scraping |
|  | Assess Gathered Information | Automating the assessment and aggregation of information from multiple sources. | AI Automation |
| Preparation | Combination and Analysis of Gathered Information | Processing and analyzing collected data to identify patterns and vulnerabilities. | AI Analysis |
|  | Development of an Attack Vector | Crafting personalized attack vectors, like phishing emails, based on gathered information. | Generative AI |
| Develop Relationship | Rapport Building | Engaging in conversations to build rapport with the target. | AI Chatbot |
|  | Establishment of Communication | Automating the initiation and maintenance of communication with the target. | AI-Automation |
| Exploit Relationship | Prime the Target | Generating content aligned with the established relationship to manipulate responses. | Generative AI |
|  | Elicitation | Employing psychological techniques through AI chatbots to elicit information or actions. | AI Chatbot |
| Debrief | Maintenance | Automating periodic interactions to sustain the established relationship. | AI Automation |
|  | Transition | Facilitating the smooth handover of interactions from one attacker to another. | AI Coordination |
| Goal Satisfaction | Tracking and Analysis | Tracking and analyzing the success of the attack in achieving its goals. | AI Assessment |

## 4.1 Realistic Content Generation

The easiest identifiable potential of Gen AI is the creation of realistic content. A good example – especially in the context of phishing – would be website cloning. AI can rapidly clone legitimate websites and modify them subtly to deceive victims, leading to more effective phishing pages. More concerningly, generative AI is gradually mastering different capabilities in text and media generation:

- Text: It can craft convincing emails, text messages, or social media posts that appear to be written by humans. These messages may be customized to target specific individuals or organizations.
- Images: It can create realistic images, such as fake IDs, official logos, or manipulated photographs, to enhance the credibility of phishing or social engineering attacks.
- Voice: It can produce realistic voice recordings or engage in voice phishing (vishing) by impersonating trusted individuals or organizations over the phone.
- Videos: It can create highly realistic deepfake videos that manipulate the appearance and voice of real individuals. These deepfakes can be used to spread misinformation, impersonate executives or public figures, and deceive targets into taking specific actions [26].

These manipulated media appear highly realistic and can deceive viewers into believing false narratives or events. For example, Deepfakes have raised significant concerns due to their potential to spread misinformation and undermine trust in visual and auditory evidence. In the context of SE and phishing, deepfakes can be created from datasets of personal visual and audio information. These datasets can be gathered from manual or automated open-source intelligence (OSINT), close-source information (CSINT), or even shared by other attackers. The combination of increasingly available personal information and increasingly sophisticated deepfake capability can be used for the following purpose:

- Impersonation: By creating a near-flawless visual and auditory imitation of a trusted individual - such as colleagues, friends, or family members - malicious actors can convince recipients to carry out tasks or disclose sensitive information.
- Bypassing Biometrics: Some security systems that rely on facial or voice recognition could be tricked by a well-crafted deepfake, thereby granting unauthorized access.
- Ransom Campaigns: Threat actors could create scandalous or compromising deepfakes of individuals and then threaten to release them unless a ransom is paid [27].
- Disinformation: On a broader scale, deepfakes can spread misinformation or propaganda that appears genuine, thus causing panic, influencing public opinion, or even affecting stock market behavior.

The evolution of AI has not stopped with visual manipulations. Recently, the introduction of sophisticated large language models (LLMs), such as ChatGPT or Bard, introduced another layer of potential misuse in the field of SE and phishing. LLMs can craft highly convincing and personalized phishing emails that are contextually relevant to the recipient, vastly increasing the success rate of such attacks. This is of course also true for all other text-based communication channels, such as SMS or Social Media Chats [17]. Instead of static phishing messages, cybercriminals could leverage LLMs to conduct real-time interactions with victims, adapting the conversation dynamically to manipulate the target more effectively. In addition, those models are developing very fast and gradually gaining new capabilities beyond text generation (i.e., ChatGPT Code Interpreter). An emerging concern with advanced AI systems like LLMs is the potential for "jailbreaking" — bypassing imposed usage restrictions or guidelines. Malicious actors, by finding and exploiting vulnerabilities in interfaces or access controls, can used AI capabilities beyond their originally intended purpose [12,28,29]. In the context of SE and phishing, such unrestricted access can supercharge deceptive campaigns, making them highly sophisticated and more challenging to identify and counter.

## 4.2 Advanced Targeting and Personalization

Generative AI has the potential to start an era of hyper-personalized and highly targeted attacks. This is because of the introduced ability to generate realistic content (pretexting), but also due to the potential of AI to analyze existing information (reconnaissance), which allows attackers to customize their malicious intents according to the online presence, behavior, and affiliations of the targets. In the final step, this information can be successfully leveraged during the execution stage.

The primary goal of reconnaissance (information gathering and intelligence) is to collect information about the target, which would allow the attacker to craft a more effective and tailored attack strategy in the subsequent stages. With its mastery of language and analytical abilities, Generative AI can scrutinize the digital footprints of targets. This provides insights into a target's specific interests, affiliations, or behaviors. Such insights could pave the way for crafting deeply personalized phishing content or social engineering attacks. This gathered intelligence can subsequently be used to develop the attack strategy – referred to as pretexting. Pretexting is a broad stage that encapsulates the creation of a story, scenario, or identity that an attacker uses to engage with the target. Given the vast capabilities of Gen-AI, it can significantly enhance this phase by producing realistic and tailored content, such as texts, images, voices, and videos (Section 3.1). This enables context-aware phishing, where AI crafts malicious content that resonates with the target's communication patterns, making the story or scenario highly believable. This might include emails that sound like they're from colleagues, friends, or familiar institutions. During execution, AI exploits identified vulnerabilities and uses generated synthetic content to manipulate victims and achieve the objectives set during the reconnaissance and pretexting phases. A good example is automated spear phishing, where AI uses those insights to generate and distribute personalized fraudulent emails or messages. Also, intelligent chatbots can ask leading questions to extract sensitive information or directly execute certain actions. Smart bots, such as "jailbroken" LLMs (i.e., ChatGPT and Bard) have shown superior capabilities to pass the Turing test, and in general, appear like "real" humans [30]. This is especially the case during purely text-based conversations.

### 4.3 Automated Attack Infrastructure

The last pillar is automation. AI can help to automate the creation and dissemination of deceptive content, thereby enabling attackers to carry out large-scale phishing campaigns or social engineering attacks with minimal effort. As established in section 3.2, AI algorithms can be used to learn about the victim's behaviors and predict their responses, which can make phishing attempts more convincing and successful. LLMs could be integrated into broader SE toolkits, offering automated and intelligent capabilities, like crafting believable responses or guiding the attacker based on real-time data. By introducing reconnaissance feedback loops at both the pretexting and execution stages, the AI becomes a dynamic system, capable of evolving its methods based on real-world outcomes. For instance, if an AI-driven chatbot notes hesitation or suspicion in a victim's responses, it could change its approach, tone, or the information it provides to appear more legitimate. This makes it a formidable tool in the hands of malicious actors and underscores the importance of robust countermeasures. The ultimate threat is a completely autonomous intelligent social engineering bot.

This brings us the next point, which is the large-scale distribution of such bots. AI significantly escalates the potential for expansive and sophisticated campaigns. Through AI-powered botnets, attackers can rapidly craft and disseminate personalized phishing messages to a massive audience. These botnets, equipped with AI capabilities, not only automate the distribution process but can also adapt the content in real time based on user interactions and feedback. Moreover, the integration of intelligent bots within these networks allows for autonomous responses to potential victims, convincingly simulating genuine human engagements. This confluence of automation, adaptability, and vast reach introduced by AI presents a heightened challenge, necessitating advanced countermeasures in our cybersecurity frameworks.

Furthermore, generative AI can generate content that evades detection by security software or filters, such as crafting phishing emails that avoid common red flags or creating malware that bypasses antivirus software. AI can use polymorphic attack patterns (e.g., emails that constantly change, dynamically altered malicious websites, etc.). Those AI-created evasion and camouflage techniques could make it very difficult to filter out malicious content. Similar to feedback loops, adapting attack models, can learn from failed attempts, continually improving their methods and becoming more effective at evading detection and/or – as mentioned earlier – deceiving their targets. Overall, using automation, attackers can execute social engineering and phishing campaigns at a scale and speed that would be impossible for human actors, significantly increasing the effectiveness and potential damage of those attacks.

### 5 GENERATIVE AI SOCIAL ENGINEERING FRAMEWORK

The Generative AI Social Engineering (GenAI-SE) Framework (see Figure 3) is a flexible and multi-dimensional model designed to investigate the complex nature of AI-driven social engineering (SE) attacks. On the one hand, the framework serves as an adaptable threat intelligence template that accommodates a diverse set of layers or dimensions for analysis and helps to evaluate a prior the impact of AI capabilities on existing and emerging threats. Hence, the framework can be used to inform decision-makers, security professionals, and automated security systems about risks and vulnerabilities so that proactive measures can be taken to improve security and mitigate risks. On the other hand, it can be used for further research to investigate different aspects of AI-driven social engineering attacks.

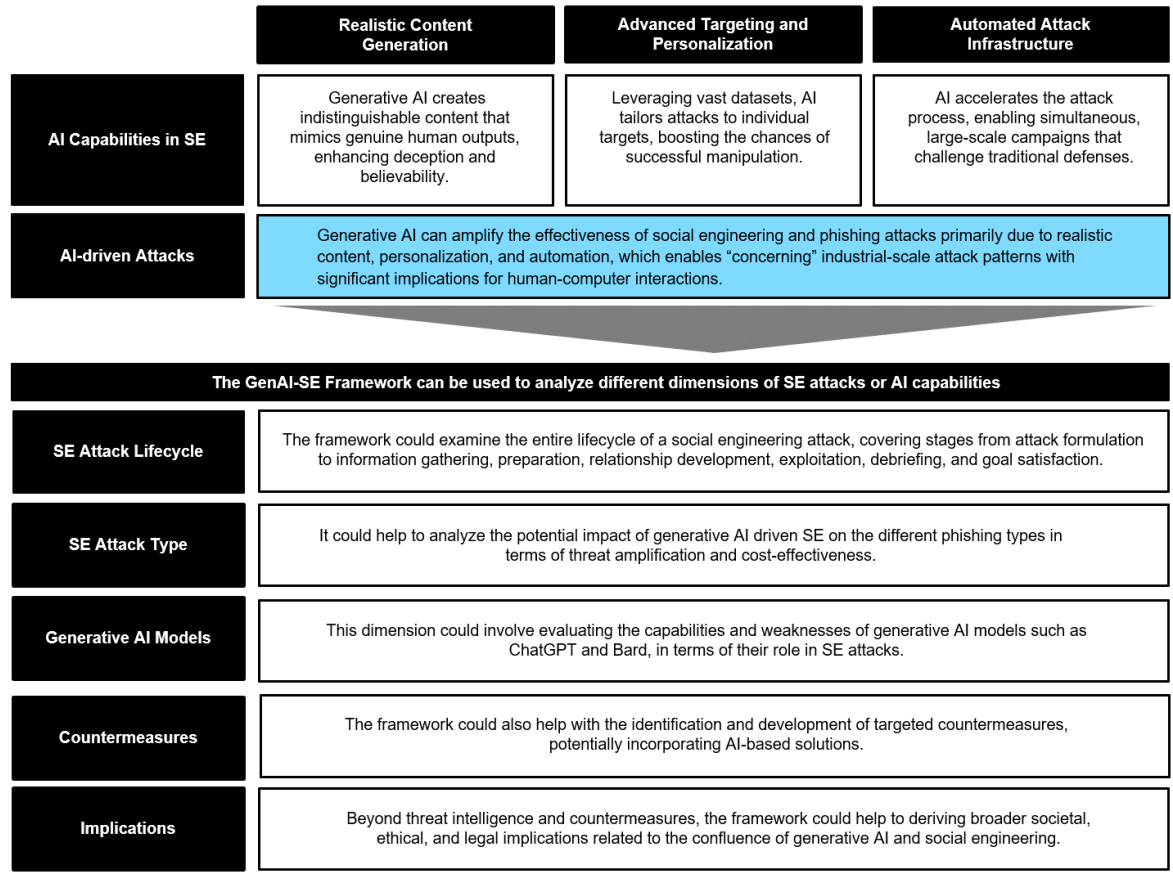

Figure 3: Generative AI Social Engineering (GenAI-SE) Framework

## 5.1 Threats Amplification and Cost-Effectiveness

The developed framework helps to evaluate a priori the implications of AI capabilities on existing and emerging threats. Figure 4 shows an analysis of the potential of AI to amplify phishing threats while at the same time reducing their costs.

| AI Capabilities in SE | Realistic Content Generation | | Advanced Targeting and Personalization | | Automated Attack Infrastructure | |
|---|---|---|---|---|---|---|
| | Generative AI creates indistinguishable content that mimics genuine human outputs, enhancing deception and believability. | | Leveraging vast datasets, AI tailors attacks to individual targets, boosting the chances of successful manipulation. | | AI accelerates the attack process, enabling simultaneous, large-scale campaigns that challenge traditional defenses. | |
| Evaluation Metric | Threat Amplification | Cost Effectiveness | Threat Amplification | Cost Effectiveness | Threat Amplification | Cost Effectiveness |
| Spam Phishing | +++ | +++ | ++ | + | +++ | +++ |
| Spear Phishing | ++ | ++ | +++ | +++ | ++ | ++ |
| Whaling | ++ | ++ | +++ | +++ | + | + |
| Interpretation | Generative AI can amplify the effectiveness of social engineering and phishing attacks primarily due to realistic content, personalization, and automation, which enables "concerning" industrial-scale attack patterns with significant implications for human-computer interactions. | | | | | |

Figure 4: Evaluation of threat amplification and cost-effectiveness of Gen-AI in SE

Advancements in AI have a significant impact on the context of phishing attacks, favoring the attacker rather than the defender. This dual nature of AI presents important implications for the security landscape. Traditional approaches that rely on training end-users to identify deception techniques are insufficient, particularly in the case of targeted attacks such as spear phishing and whaling. Humans often struggle to detect and defend against these sophisticated techniques, leaving organizations vulnerable to highly personalized and convincing phishing campaigns. The use of sophisticated AI holds the potential to further improve tailored attacks for the mass market. Attackers can leverage AI algorithms to analyze large datasets and create highly targeted phishing campaigns at scale. This automation and personalization significantly increase the effectiveness of spear phishing campaigns, posing a greater threat to individuals and organizations. The implications of AI-powered phishing attacks are further amplified by the increasingly sophisticated underground ecosystem of cybercriminals, the emergence of the gig economy, and the widespread adoption of Software as a Service (SaaS) solutions. This underground ecosystem thrives on the exchange of tools, techniques, and resources, providing attackers with a fertile ground to leverage AI for their malicious activities. The collaboration and sharing of AI-powered attack methods within this ecosystem pose a significant challenge for defenders. Additionally, the gig economy introduces new complexities by offering accessible platforms that facilitate the utilization of AI technologies, tools, and services. This ease of access empowers even non-expert individuals to employ AI for nefarious purposes, further contributing to the prevalence of AI-enabled phishing attacks. Moreover, the adoption of SaaS solutions provides attackers with readily available AI-powered tools and infrastructure to launch sophisticated phishing campaigns.

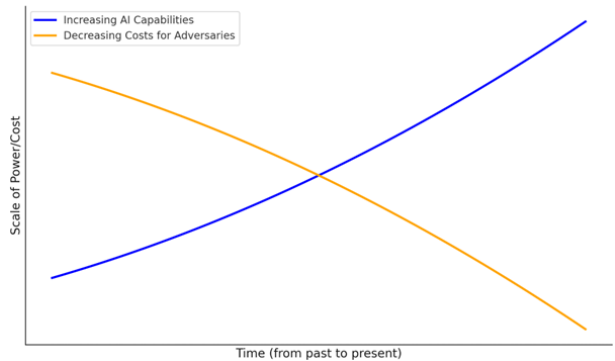

Figure 5: Relationship Between AI Capabilities and Costs

Furthermore, the continuing trend of AI becoming more cost-effective is a double-edged sword in this context. While these financial efficiencies benefit many industries by making AI more accessible and democratized, they simultaneously lower the barrier to entry for malicious actors. As the cost of acquiring and deploying AI capabilities decreases (see Figure 5), a broader array of adversaries, even those with limited resources, can tap into sophisticated AI tools for malicious purposes. This means that the scale and frequency of AI-driven phishing attacks could surge, making it even more challenging for organizations to maintain an effective defense. The interplay of accessibility and affordability of AI technologies may inadvertently intensify the phishing threat landscape, underscoring the urgency for novel defensive measures.

### 5.2 Countermeasures for AI-Generated Deception

The nature of cybercrime as a continuously evolving threat—tug of war between adversaries and defenders—makes it impossible to determine guaranteed protection measures [7]. On the highest level we can distinguish between technical countermeasures, which are largely based on AI/ML [31] and cryptography, and human-centric countermeasures, which is largely user education [32]. While ideas for innovative solutions exist [33–35], the cybersecurity industry is focusing predominantly on awareness training programs to tackle the so-called "people" problem. These programs are commonplace and applied by all types of businesses ranging from small to global corporations. Yet, they are generally ineffective against advanced targeting and personalization tactics. Labeling the end-user as the problem is not just an oversimplification, but a dangerous perspective. The implication is that users should be blamed for not being smart/aware enough and remediate this through security awareness, education and training (SAET), however this approach is insufficient and flawed for a number of reasons. (1) In the face of AI improvements in mimicking human interactions, it is not feasible for this increasing capability to be mitigated through greater awareness or understanding. We are fast approaching a point where people will not be able to differentiate genuine from fabricated content. (2) The potential scale and automation of malicious attacks will mean that individuals will need constant and near-perfect levels of awareness to reach a satisfactory balance between protecting themselves from deception while maintaining trust in communication channels. Finally, (3) it is already difficult to evaluate the effectiveness of SAET: if someone fails a phishing test or becomes a victim of social engineering, this is usually interpreted as a sign that they weren't trained enough, or paying enough attention, which is then used as justification for more awareness, training and education. This is circular reasoning, which buys into the harmful fallacy that victims are to be blamed for falling prey to attackers.

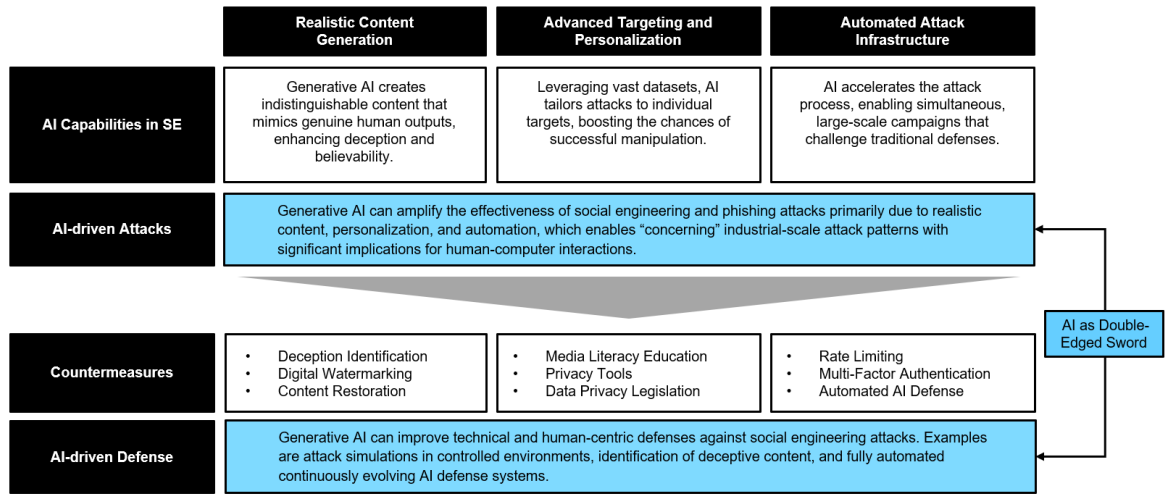

Figure 6: Derivation of countermeasures for AI-generated deception

Even seasoned cybersecurity experts are susceptible to sophisticated AI-generated phishing attacks. The expectation that individuals can scrutinize every email they receive is impractical and counterproductive, especially in today's fast-paced digital environment, that comes with heavy cognitive workload [36]. Also, this mindset ignores essential psychological factors explained by researchers like Daniel Kahneman [37]. Routine tasks, like checking emails, seldom involve critical thinking, and when emotions—especially fear—are triggered, rational decision-making becomes challenging. Current solutions like anomaly and spam detection AI are somewhat effective, but they fall short against targeted spear phishing and state-sponsored attacks.

The implications of AI in the context of phishing highlight the need for innovative defenses and proactive strategies to mitigate the risks posed by AI-enabled attacks. Table 6 contains a number of indicative proposals for specific countermeasures for each of the three pillars. It is important to note that while generative AI can be used maliciously in these scenarios, it can also be employed for ethical purposes, such as creating training materials for cybersecurity professionals or developing tools to detect and counter AI-generated deceptive content. Overall, the relentless evolution of AI makes it paramount to stay updated on the latest advancements in AI-driven deception and its countermeasures. In mapping AI capabilities to the detail of social engineering attacks, we demonstrate how the framework can be used to highlight new areas of development and research in disrupting AI powered social engineering attacks.

Table 6: Countermeasures for AI-Generated Deception

| Category | Countermeasure | Explanation |
| --- | --- | --- |
| Realistic Content Creation | Deception Identification/AI Verification Tools | Develop AI-based solutions that can recognize and flag AI-generated content. For instance, tools that detect deepfakes by analyzing inconsistencies in videos. |
| | Digital Watermarking | Implement watermarking techniques for genuine content, signaling to users the authenticity of a digital item. |
| | Content Restoration | When content undergoes malicious AI modification, generative AI models can attempt to revert or rectify the content, such as reversing deepfake alterations. |

| Category | Countermeasure | Explanation |
|---|---|---|
| Advanced Targeting and Personalization | Media Literacy Education | Educate the public about the capabilities of generative AI to put them in a position to recognize potential AI-generated content. |
| | Privacy Tools | Encourage the use of privacy tools that block tracking cookies, scripts, and other methods used to gather personal data for targeted attacks. |
| | Data Privacy Legislation | Strengthen regulations around data collection, storage, and usage to limit the extent of personal data available for AI targeting. |
| Automated Attack Infrastructure | Rate Limiting | Implement rate limiting for message sending or content posting on platforms, making mass phishing attempts via automation more difficult. |
| | Multi-Factor Authentication | Promote the use of MFA, making it harder for automated systems to gain unauthorized access, even if they have some user credentials. |
| | Automated Defense | Using AI-enabled defenses based on behavioral analysis, adversarial training, and simulated attacks, systems can be automatically fortified against threats. |

# 6 DIRECTIONS FOR FUTURE RESEARCH

The three identified capabilities of generative AI in the context of social engineering—realistic content, personalization, and automation—create a problematic cocktail that marks the onset of potentially unforeseeable innovations in hacking. It is easy to imagine the development of powerful and autonomous social engineering bots, and enhancing existing and developing novel countermeasures for these emerging threats becomes paramount. We recommend the following avenues for future research:

- **User Awareness and Education:** We need to recognize the fact that "awareness training" is not an easy fix against AI-powered cyber-attack. Nevertheless, increasing user knowledge about the latest techniques and strategies can empower individuals to recognize and respond appropriately to potential threats [38,39]. This could involve designing effective training programs, interactive simulations, and user-friendly educational materials [34,35].
- **Adversarial Machine Learning:** Develop advanced techniques to detect and defend against adversarial attacks in the context of AI-powered social engineering. Adversarial machine learning aims to create robust models that can withstand manipulation attempts by attackers. Research in this area can focus on developing algorithms and strategies to detect and mitigate adversarial SE and phishing techniques.
- **Active Deception Defense:** Develop proactive defense mechanisms that actively disrupt deceptive attacks in real-time. This could involve technologies like natural language processing, anomaly detection, and real-time analysis of communication channels to identify and block SE and phishing attempts as they occur.
- **Explainable AI for Threat Detection:** Enhance the transparency of AI models used for threat detection. Explainable AI techniques can help security analysts and users understand how AI systems make decisions and identify indicators of deceptive attacks. By providing meaningful explanations and insights, it becomes easier to trust and validate the outputs of AI-powered threat detection systems [40].
- **Emerging Technologies:** Future research could explore the intersection of emerging technologies and their implications for social engineering attacks. As technologies, such as Chatbots [41], brain-computer interfaces [42], robotics [43,44], quantum computing [45], and metaverse ecosystems [46,47] become more integrated into our daily life, new avenues for social engineering attacks are likely to emerge.

## 7 CONCLUSION

The potential for persistent high-volume social engineering (SE) attacks poses a serious threat to the trust that users place in human-computer interactions (HCI). When these interactions are compromised through continual SE attacks, whether driven by generative AI or other methods, it erodes the foundation of trust and security that is essential for effective HCI. Users may become increasingly skeptical of the systems they interact with, questioning the authenticity of dialogues and hesitating to share information or perform tasks. This paper analyzed the transformative potential of Generative AI in the realm of social engineering and phishing. We argued that the integration of Generative AI can profoundly amplify the effectiveness of these cyberattacks. The primary drivers behind this enhancement are the capabilities of AI to produce realistic content, tailor-made personalization, and unparalleled automation. This potent combination paves the way for industrial-scale attack patterns, signaling a paradigm shift to a new, more sophisticated era of cyber threats. Further compounding this challenge is the evolving landscape of AI. As artificial intelligence continues its trajectory of becoming simultaneously more powerful and economically accessible, the barriers to misuse diminish. This raises an alarm for not only the cybersecurity community but also industries and individuals at large. In the face of such emerging threats, there is an acute and pressing need to develop and refine countermeasures. A proactive stance today is our best defense against the impending wave of AI-powered cyber threats of tomorrow.